\documentclass[aps,epsfig,twocolumn]{revtex4}
\usepackage{graphicx}
\usepackage{amsmath}
\usepackage{latexsym}
\usepackage{amsfonts}
\usepackage{amssymb}
\usepackage{array}
\usepackage{color}
\usepackage{subfigure}
\usepackage{verbatim}
\usepackage{dsfont}
\newcommand{\ket}[1]{\left\vert#1\right\rangle}
\newcommand{\bra}[1]{\left\langle#1\right\vert}

\begin{document}
\newcommand{\Q}[1]{{\color{red}#1}}
\newcommand{\blue}[1]{{\color{blue}#1}}
\newcommand{\red}[1]{{\color{red}#1}}
\newcommand{\Change}[1]{{\color{green}#1}}

\title{Deterministic nonlinear phase gates induced by a single qubit}

\author{
Kimin Park*, Petr Marek and Radim Filip}
\affiliation{
Department of Optics, Palack\'y University, 17. listopadu 1192/12, 77146 Olomouc, Czech Republic}
\date{\today}
\email{park@optics.upol.cz}

\begin{abstract}
We propose deterministic realizations of nonlinear phase gates by repeating non-commuting Rabi interactions feasible between a harmonic oscillator and {\em only} a single  two-level ancillary qubit. We show explicitly that the key nonclassical features of the states after the ideal cubic phase gate and the quartic phase gate are reproduced faithfully by the engineered operators. This theoretical proposal completes the universal set of operators in continuous variable quantum computation.
\end{abstract}

\maketitle

Quantum linear harmonic oscillators, systems with states defined in an infinite-dimensional Fock space, are at the center of continuous variables (CV) quantum information processing \cite{CVqinf}. Physical realizations of quantum harmonic oscillators employ a traveling light \cite{CVlight} and collective modes of atomic spins \cite{CVatoms}, motion modes of trapped ions \cite{CVions}, and cavity \cite{CVcavQED} or circuit quantum electrodynamics  \cite{CVcirQED}. In order to fully control these systems as a quantum computation platform we need the ability to implement an arbitrary nonlinear operation. The principal requirement is reduced to an access to nonlinearity of the third order - the cubic nonlinearity - and then use it for implementation of nonlinearities of higher order \cite{LloydPRL1999Universal}. However, obtaining even the cubic nonlinearity is not straightforward. The evolution in cubic potential is inherently instable, even in the overdamped regime~\cite{cubicTrapped}. Naturally appearing nonlinear media do not come with sufficient strength and purity \cite{nonlinearity} and the evolution operator needs to be crafted by manipulating individual quanta of the harmonic oscillator. These approaches rely on nonlinear conditional measurements \cite{ParkPRA2016Rabiconditional,ParkPRA2016JC,OptCub}, deterministic feed-forward aided by nonlinear ancillary states \cite{cubicstate,BartlettPRA2002cubicstate,MarekPRA2011Cubic,MiyataPRA2016Cubic}, or deterministic control over a two-level system coupled to the oscillator \cite{KrastanovConstructivePRA2015, ParkKerrSciRep2017}.

These methods are specialized to different CV systems and resources. By far the most effective approaches in quantum optics and quantum electrodynamics (QED) exploit the coupling between the harmonic oscillator and an associated two level system, which is the standard scenario in trapped ions and circuit QED. This efficiency originates from the inherent nonlinearity of the two-level system, which can impart nonclassical behavior to the oscillator simply by ``being there'' \cite{Nonclassicality}. There are two distinct approaches towards realizing the nonlinear operations, either in a single involved step \cite{KrastanovConstructivePRA2015}, or by incremental construction from weak elementary interactions \cite{OptCub,ParkKerrSciRep2017}. 
In the following,  we will focus on the method of elementary gates \cite{ParkKerrSciRep2017}, directly extended towards continuous-variable phase gates by designing a different coupling. 
Such gates offer the ability to address the wave-like features of quantum systems and can be employed for simulation of particles in nonlinear potentials \cite{simulation}.

In this Letter we propose deterministic methods for realizing a unitary nonlinear quadrature phase gate of an arbitrary order for quantum harmonic oscillators. The gates realized by our method gradually approach unitary evolutions, which take advantage of the so-called Rabi interaction between the harmonic oscillator and a coupled two-level system. They consist of ordinary elementary building blocks  simply repeated to achieve nonlinearities of arbitrary strength in the limit of infinite resources. We analyze the performance of these methods under the assumption of finite resources.

\begin{figure}[th]
\includegraphics[width=250px]{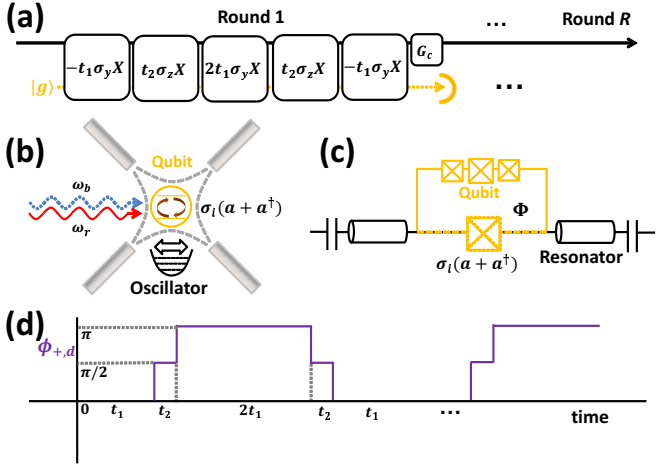}
\caption{(Color online) A deterministic setup to achieve a nonlinear phase gate. (a) A five-element circuit model where non-commuting Rabi interactions at different strengths are arranged together with a correction operator. An ancillary qubit (yellow elements) is prepared in a ground state at each round, and $R$ rounds are performed for a high-strength nonlinear gate. (b) A trapped ion implementation in a Paul trap. Here the continuous system is the motional state of
the ion, and the two-level system is its internal energy state. The
preparation and measurement of the two-level system
are performed by control pulses. Rabi interaction is realized by a bichromatic laser driving. (c) A circuit QED implementation. The flux qubit is made of a superposition of the clockwise and counter-clockwise currents, and is interacting via Josephson junction with the superconducting microwave coplanar waveguide resonator. External magnetic flux $\Phi$ controls the interaction. (d) In trapped ionic system, the phase of the bichromatic laser beam $\phi_+=(\phi_r+\phi_b+\pi)/2$ for the phase of each beam $\phi_{r,b}$ is controlled in time segments alternatingly corresponding to the assigned Pauli operators for a given strength~\cite{RabiOperationIon}. In circuit QED system, exactly the same control sequence can be obtained by controlling the phase of the external driving fields $\phi_d$ \cite{CQEDtheory}. }
\label{Setup}
\end{figure}

The continuous-variable phase gates are represented by unitary operators $\hat{U}_m=\exp[i \chi_m \hat{X}^m]$, where the arbitrary integer $m$ denotes the order of the operation. $\hat{X}=(\hat{a}+\hat{a}^\dagger)/\sqrt{2}$ is a quadrature operator of the harmonic oscillator, and $\hat{a}$ and $\hat{a}^{\dag}$ denote the ladder operators. The two-level system, or a qubit, is described by a Bloch sphere with the SU[2] algebra represented by Pauli matrices $\sigma_i$ with $i=x,y,z$ satisfying a set of commutation relations $[\sigma_i,\sigma_j]=2i\epsilon_{ijk}\sigma_k$ with a Levi-Civita symbol $\epsilon_{ijk}$. The main ingredient is a  sequence of individual Rabi interactions between the harmonic oscillator and two level systems, already feasible at the various platforms  \cite{RabiOperationCQED,RabiOperationIon}. A coupling referred to as the Rabi interaction \cite{RabiHist} is represented by a unitary operator $\exp[it\sigma_i \hat{X}]$, where $t$ is its effective strength. In trapped ionic system, this interaction can occur between a motional degree of freedom and the atomic internal two-level states via bichromatic laser driving red and blue motional sidebands in Lamb-Dicke regime~\cite{RabiOperationIon}. In circuit QED system, this type of interaction arises between a superconducting qubit such as a flux or a transmon qubit and a waveguide resonator in the ultrastrong-coupling regime~\cite{RabiOperationCQED}. 

Rabi gates with different orientation of  $\sigma_i$ do not commute and we can exploit this non-commutative behavior to elaborate important ingredients before the proposal of phase gates. With the help of the Baker-Campbell-Hausdorff formula~\cite{BCH}, new combined operators to be used as building blocks are derived as:
\begin{align}
\hat{M}_1(\hat{T}_1,\hat{T}_2)&=\exp[-i \hat{T}_1 \sigma_y] \exp[i \hat{T}_2 \sigma_z]\exp[i\hat{T}_1 \sigma_y]\nonumber\\
&=\exp[i\hat{T}_2(\cos[2\hat{T}_1]\sigma_z-\sin[2\hat{T}_1]\sigma_x)],\nonumber\\
\hat{M}_2(\hat{T}_1,\hat{T}_2)&=\exp[-i \hat{T}_1 \sigma_y] \exp[i \hat{T}_2 \sigma_x]\exp[i\hat{T}_1 \sigma_y]\nonumber\\
&=\exp[i\hat{T}_2(\cos[2\hat{T}_1]\sigma_x+\sin[2\hat{T}_1]\sigma_z)],
\label{eq:commutation}
\end{align}
where the harmonic oscillator operators $\hat{T}_{1,2}$ introduced for a shorthand notation commute with all the other operators in the formula. 
When the qubit is prepared in the ground state $\ket{g}$ of an eigenstate of $\sigma_z$, subject to a sequence of two combined operators $\hat{M}_{1,2}$, and subsequently projected onto the ground state by a suitable measurement, the resulting operator is expressed as
\begin{align}
&\hat{O}_{3,s}\equiv\bra{g}\hat{M}_1(\hat{T}_1,\hat{T}_2)\hat{M}_1(-\hat{T}_1,\hat{T}_2)\ket{g}\approx\exp[i \hat{T}_2\cos[2\hat{T}_1]],\nonumber\\
&\hat{O}_{4,s}\equiv \bra{g}\hat{M}_2(\hat{T}_1,\hat{T}_2)\hat{M}_2(-\hat{T}_1,-\hat{T}_2)\ket{g}\approx \exp[i\hat{T}_2 \sin [2\hat{T}_1 ]].\label{eq:refined}
 \end{align}
 Here the approximation holds when the eigenvalues of the operators $\hat{T}_1$ and $\hat{T}_2$ relevant for the input states are all small enough \cite{ParkPRA2016Rabiconditional}.  In our case, $\hat{T}_1=t_1 \hat{X}$ and $\hat{T}_2=t_2 \hat{X}$, and the condition requires the states to be localized in X-representation. For a small $t_1$ and $t_2$ the gate is practically applicable to an arbitrary state.
The importance of these new operators $\hat{O}_{3,s}$ and $\hat{O}_{4,s}$ lies in the observation that they correspond to interaction Hamiltonians no longer linear in $\hat{T}_i$. They instead contain trigonometric functions of $\hat{T}_i$, or infinite-order polynomials containing all orders of non-linear terms. These terms can be exploited in realization of the nonlinear operations of an arbitrary order.

Let us first begin with the realization of the cubic gate represented by the unitary operator $\hat{U}_3=\exp[i\chi_{3} \hat{X}^3]$ with the target strength $\chi_{3}$. We can start from (\ref{eq:refined}) with Rabi interactions arranged as in Fig.~\ref{Setup}(a) with the substution $\hat{T}_1=t_1 \hat{X}$ and $\hat{T}_2=t_2 \hat{X}$ as
  \begin{align}
&\hat{O}_{3,s}
\approx \bra{g}\exp[i2t_2\hat{X}\cos[2t_1 \hat{X}]\sigma_z]\ket{g}\nonumber\\
&=\exp[i2t_2\hat{X}\cos[2t_1 \hat{X}]].\label{eq:5-query-c}
\end{align}
The last form can be expanded into the Taylor series of only two terms  as $\exp[i2t_2\hat{X}\left(1 - (2t_1 \hat{X})^2/2\right)]$ for small values of $t_1$. This operator represents a desired cubic gate, but with an residual displacement $\hat{D}(-2 t_2)$ where $\hat{D}(z) = \exp(-i z \hat{X})$. This displacement, however, can be compensated by an inverse correction displacement operation $\hat{G}_c=\hat{D}(2 t_2)$ which can be achieved with another Rabi interaction. A weak cubic gate can therefore be obtained as a sequence of operations $\hat{G}_c\hat{O}_{3,s}$. This operator is close to a unitary operator and reliably approximates the desired operation $\hat{U}_3$ with the target strength $\chi_3 = 4 t_1^2 t_2 $ for small Rabi strengths $t_1, t_2\ll 1$ (see supplemental material for details). This near-unitarity  arises from the fact that the operator $\exp[i2t_2\hat{X}\cos[2t_1 \hat{X}]\sigma_z]$ in (\ref{eq:5-query-c}) has $\ket{g}$ as its eigenstate, beneficial in experimental implementation because the success probability of the operation approaches one. The final projective measurement on ancilla is thus of a minor contribution, and we may choose to ignore the measurement outcomes and achieve a deterministic operation represented by a trace preserving map
\begin{align}\label{deterministic}
\Gamma[\rho]=\hat{G}_c (\hat{O}_{3,s}\rho \hat{O}_{3,s}^\dagger+\hat{O}_f\rho \hat{O}_f^\dagger) \hat{G}_c^\dagger,
\end{align}
where $\hat{O}_f=\bra{e}\hat{M}_1(t_1 \hat{X},t_2 \hat{X})\hat{M}_1(-t_1 \hat{X},t_2 \hat{X})\ket{g}=-\sin ^2(t_2 \hat{X}) \sin (4 t_1  \hat{X})$ represents the failure operator vanishing for weak Rabi strengths $t_1, t_2\ll 1$. In order to enhance the total strength of the nonlinearity we may apply the weak gate multiple times.  After $R$ repetitions, the output state $\rho_\mathrm{re}=\Gamma^R[\rho_0]$ produced by the cubic gate feels the effective strength $\chi_3 = 4 t_2 t_1^2 R$. We note that two free parameters exist among $t_1$, $t_2$, and $R$ to achieve a target operator at a fixed strength $\chi_3$. These free parameters can be used to optimize the performance of the target gate.
The cancellation by the correction displacement operator $\hat{G}_c$ can equivalently be done altogether at once after all the repetitions of rounds instead at each round, due to the commutativity of $\hat{G}_c$ with $\hat{O}_{3,s}$ and $\hat{O}_f$ to simplify the experimental setup. 

Instead of using five elementary Rabi interactions as in (\ref{eq:5-query-c}), different numbers of Rabi interactions can be used. Weak cubic gate can be also obtained by using three elementary Rabi interactions $M_1(t_1 \hat{X},t_2 \hat{X})|g\rangle$, or two Rabi interactions $M_1(t_1 \hat{X},t_2)|g\rangle$ in place of $\hat{O}_{3,s}$ in (\ref{deterministic}). 
We can also effectively increase the number of Rabi interactions in each round by reducing the frequency of re-initialization of the ancillary state as $(\hat{M}_1(t_1 \hat{X},t_2\hat{X})\hat{M}_1(-t_1 \hat{X},t_2\hat{X}))^k\ket{g}$ in place of $\hat{O}_{3,s}$ for integers $k>1$.  
The performance of these approaches is, however, worse for equal numbers of elementary Rabi interactions (see supplemental material). 
 We shall therefore consider only the best operation (\ref{eq:5-query-c}) for the following analysis.

\begin{figure}[th]
\includegraphics[width=120px]{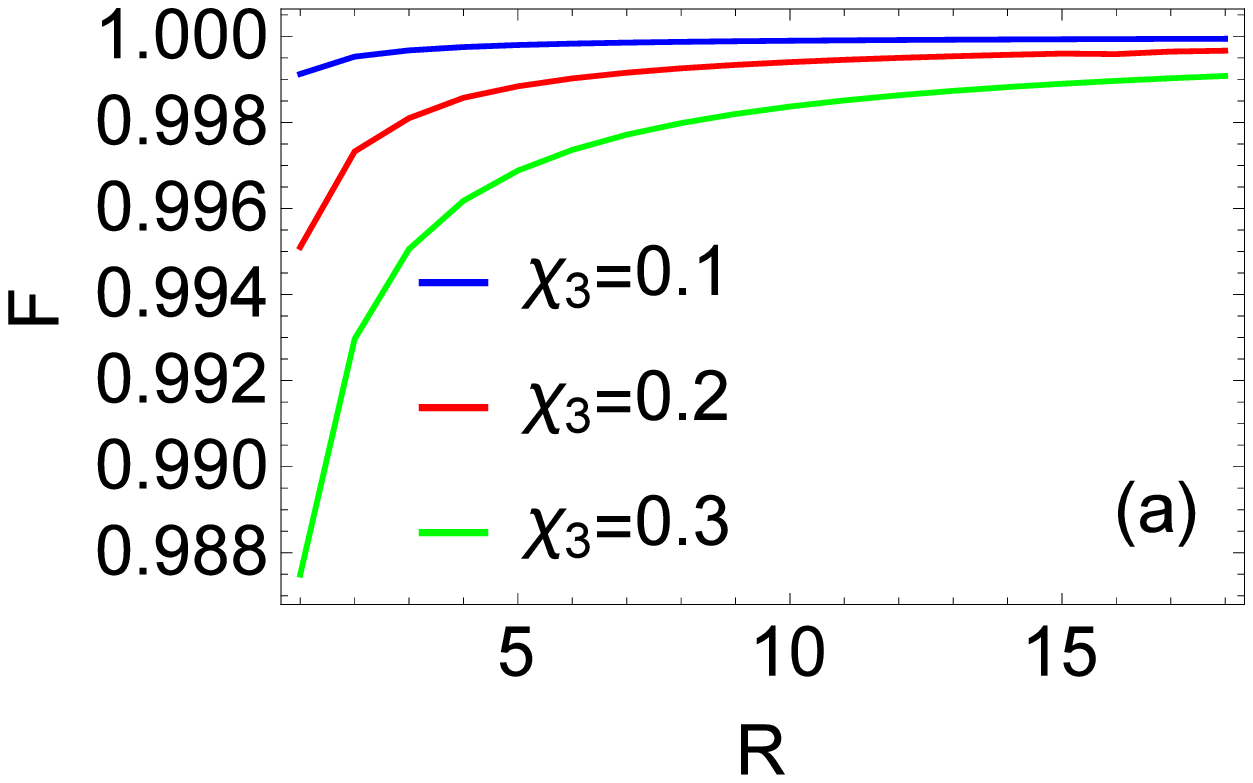}
\includegraphics[width=120px]{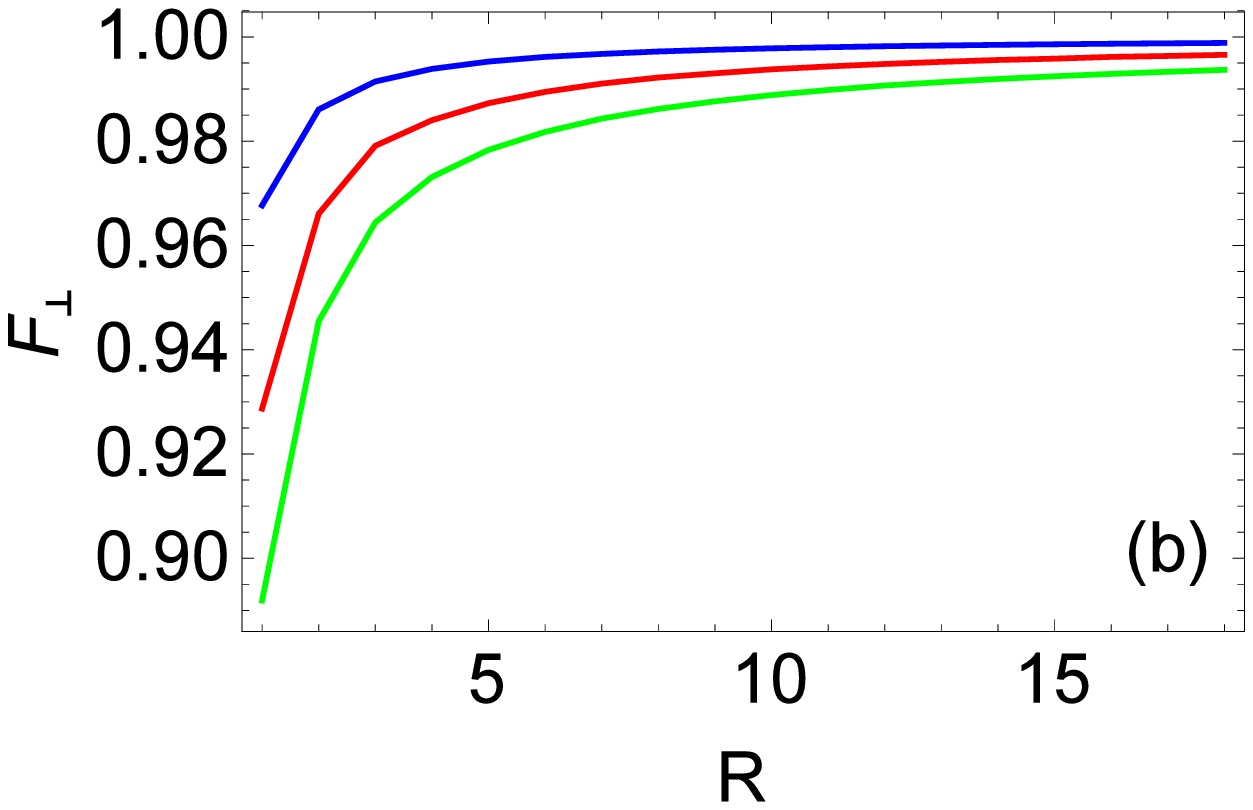}
\includegraphics[width=120px]{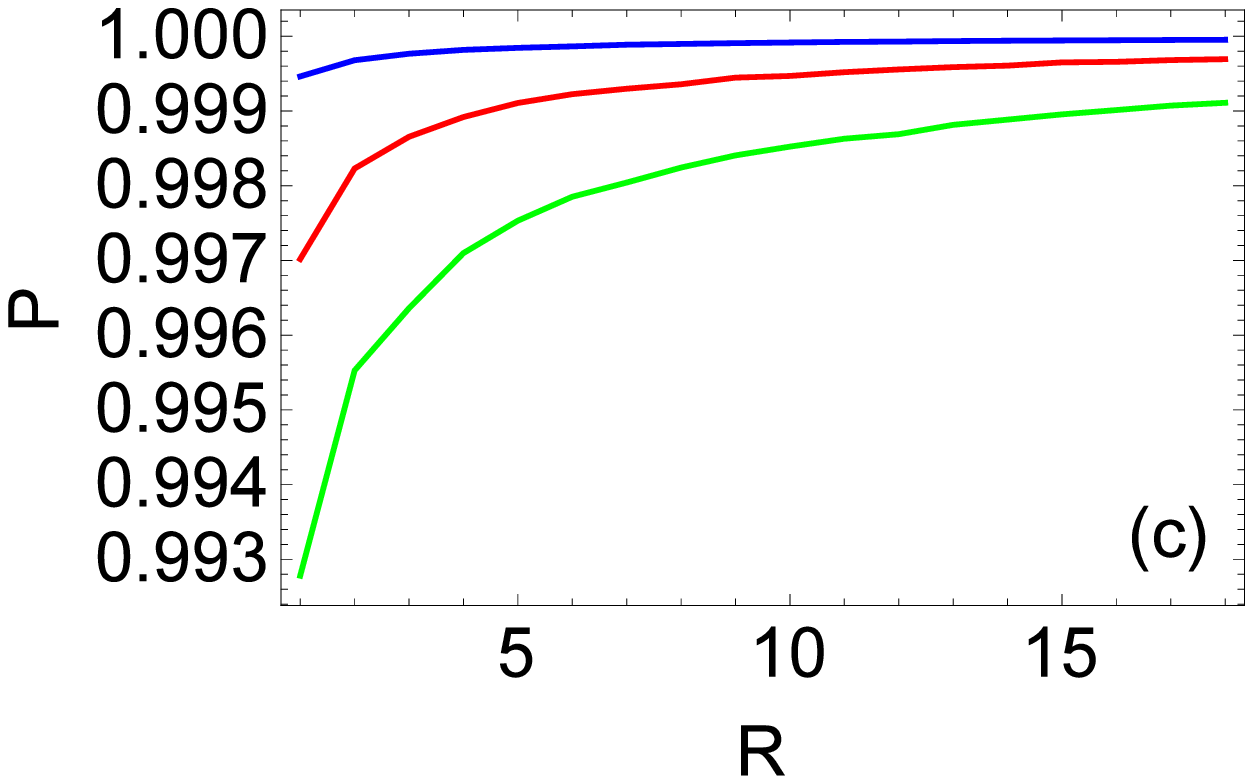}
\includegraphics[width=120px]{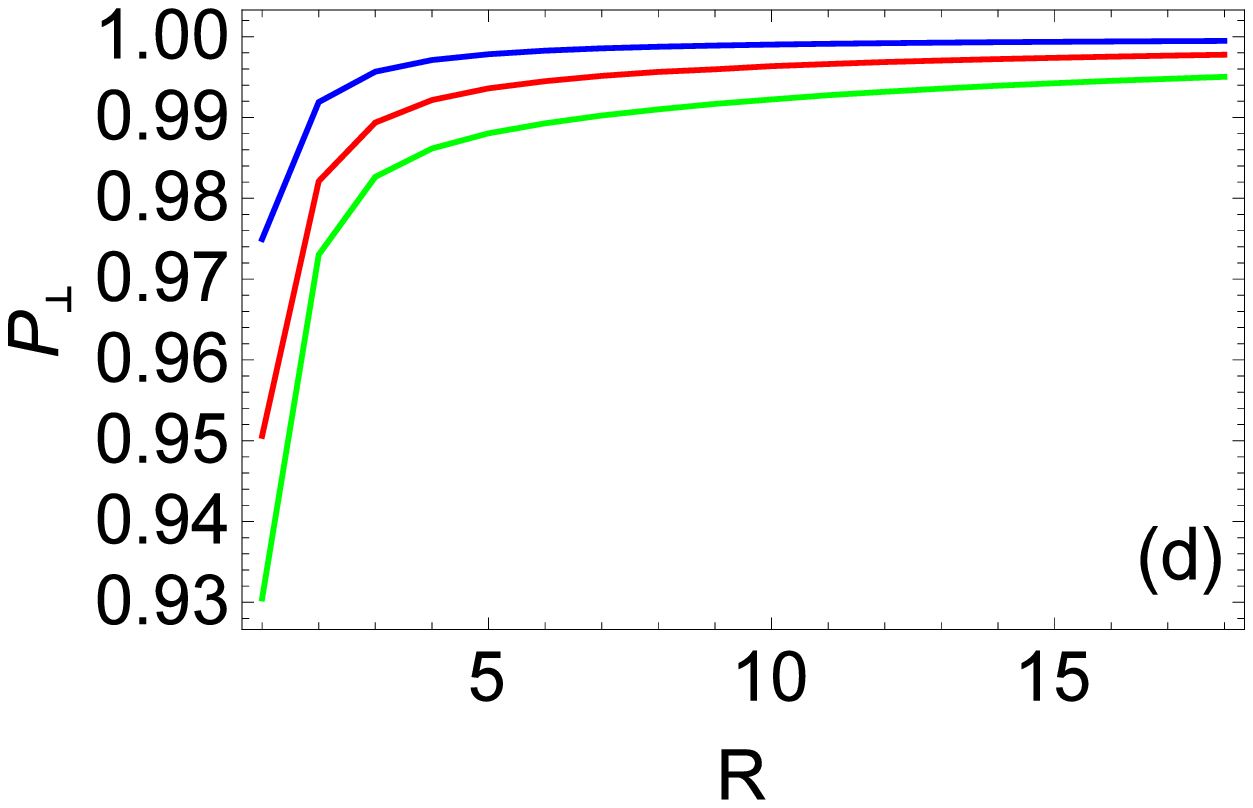}
\caption{ (Color online) Dependence of functionalities of deterministic cubic gates on the number of rounds $R$. (a) Fidelity $F$ vs $R$, (b) fidelity of change $F_\perp$ vs $R$, (c) purity $P$ vs $R$ at various strengths $\chi_3=0.1$ (blue), $0.2$ (red), $0.3$(green). All properties of the output states are enhanced as $R$ is increased for all strength $\chi_3$, and (d) purity of the orthogonal state $P_\perp$ vs $R$ at various strengths $\chi_3=0.1, 0.2, 0.3$. All properties of the output states are enhanced as $R$ is increased for all strength $\chi_3$.}
\label{repetition}
\end{figure}

Let us now evaluate the performance of the engineered cubic gate. We start by applying the operation to  an initial pure quantum state $\rho_0=\ket{\psi_0}\bra{\psi_0}$ and verify how close  the resulting state is to its ideal form. This can be quantified by the fidelity $F=\mathrm{Tr}[\rho_\mathrm{id} \rho_\mathrm{re}]$, where $\rho_\mathrm{id}=\hat{U}_3\rho_0\hat{U}_3^\dagger$ is the ideal state and $\rho_\mathrm{re}=\Gamma^R[\rho_0]$ is the realized state. In Fig.~\ref{repetition} we can see that with increasing number of repetitions corresponding to a high resource limit $R\rightarrow \infty$, the fidelity is approaching one for an arbitrary value of desired cubic interaction strength for the vacuum state $\ket{0}$. However, for small values of interaction strength, the fidelity has a weakness as a figure of merit, because it draws dominant contribution from the presence of the initial quantum state $\rho_0$ \cite{cubicstate,BartlettPRA2002cubicstate}. Therefore, the closeness of the generated state $\rho_\mathrm{re}$ to the ideal state $\rho_\mathrm{id}$ after the initial state component is removed from both states can be a complementary measure of performance without the weakness. This mathematical removal can be achieved  when a density matrix $\rho$ is projected onto a subspace orthogonal to the initial pure state $\rho_0$ as $\rho^{\perp}=(\hat{\mathds{1}}-\rho_0)\rho(\hat{\mathds{1}}-\rho_0)/\mathrm{Tr}[\rho(\hat{\mathds{1}}-\rho_0)]$. The fidelity after the initial state removal between the target state and the simulated state $F_\mathrm{\perp} = \mathrm{Tr}[\rho_\mathrm{id}^\perp \rho_\mathrm{re}^\perp]$ will be denoted as the fidelity of change, and is shown in Fig.~\ref{repetition} for various target strengths. We again observe that $F_\mathrm{\perp}\rightarrow 1$ in the high resource limit. Another measure of performance, suitable for confirmation of the unitarity of the approximative operation, is the purity $P[\rho_\mathrm{re}]=\mathrm{Tr}[\rho_\mathrm{re}^2]$ which takes a value $1$ for a pure state. The purity is reduced by the presence of the failure operators $\hat{O}_f$ in (\ref{deterministic}), but Fig.~\ref{repetition} demonstrates that this influence can be indeed vanishing as well in the high resource limit. The purity of the state after initial state removal $P[\rho^{\perp}]$ also approaches $1$ as an auxiliary proof of the closeness of the operators.


 \begin{figure}[th]
\includegraphics[width=120px]{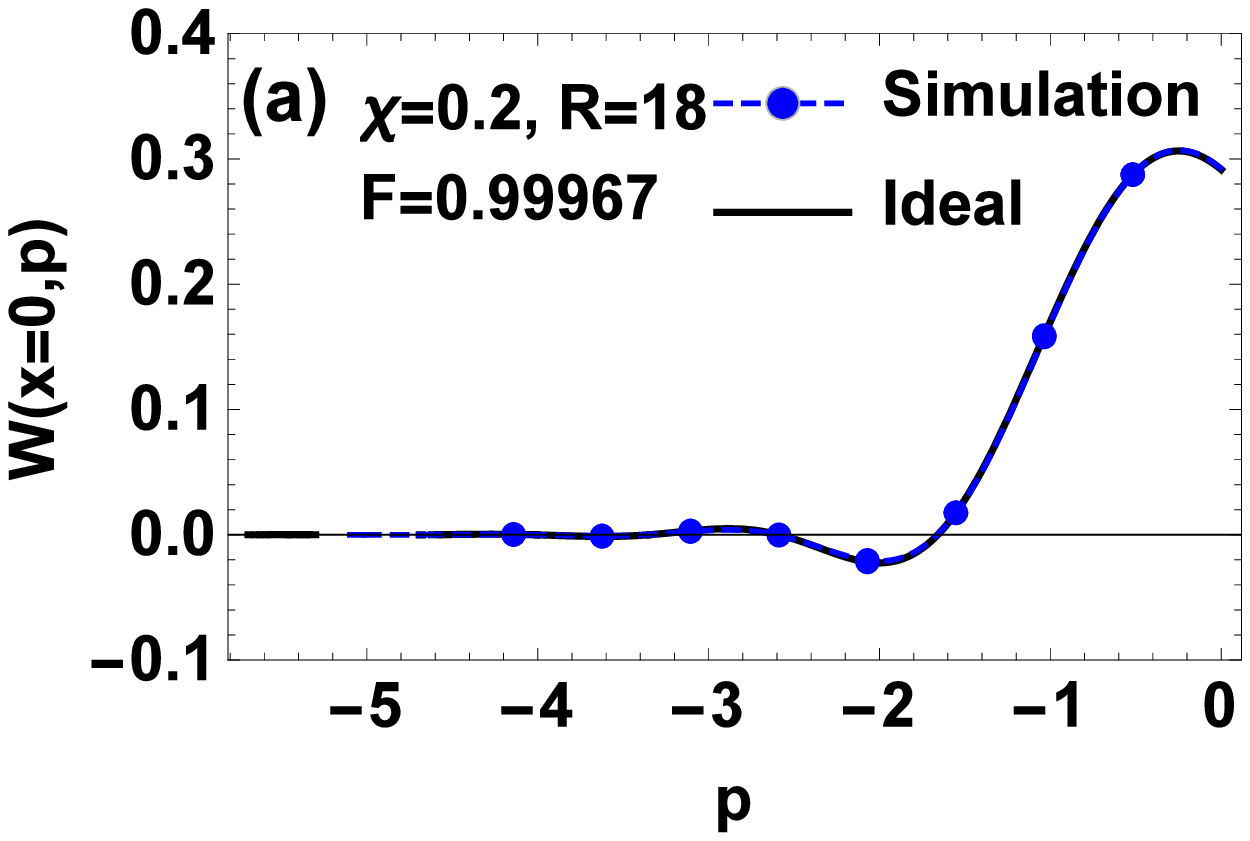}
\includegraphics[width=120px]{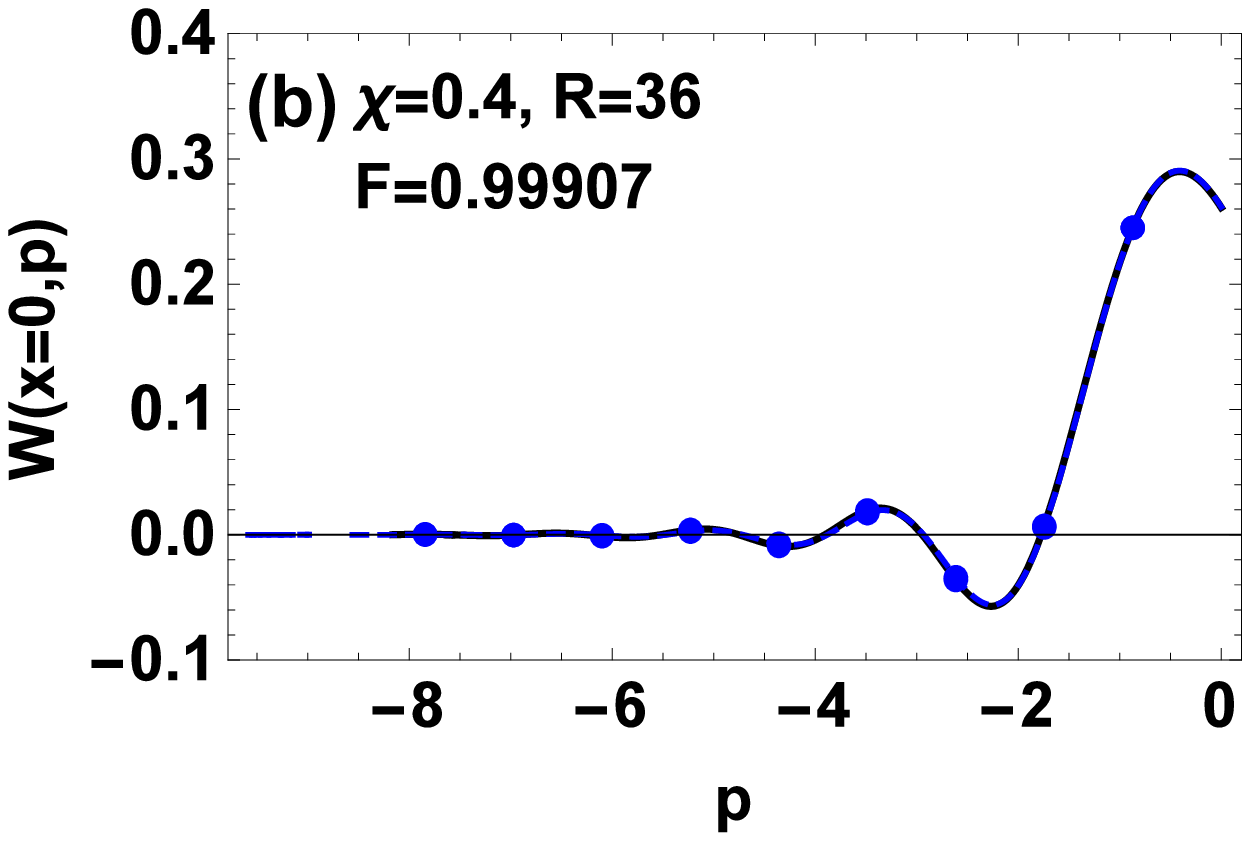}
\includegraphics[width=120px]{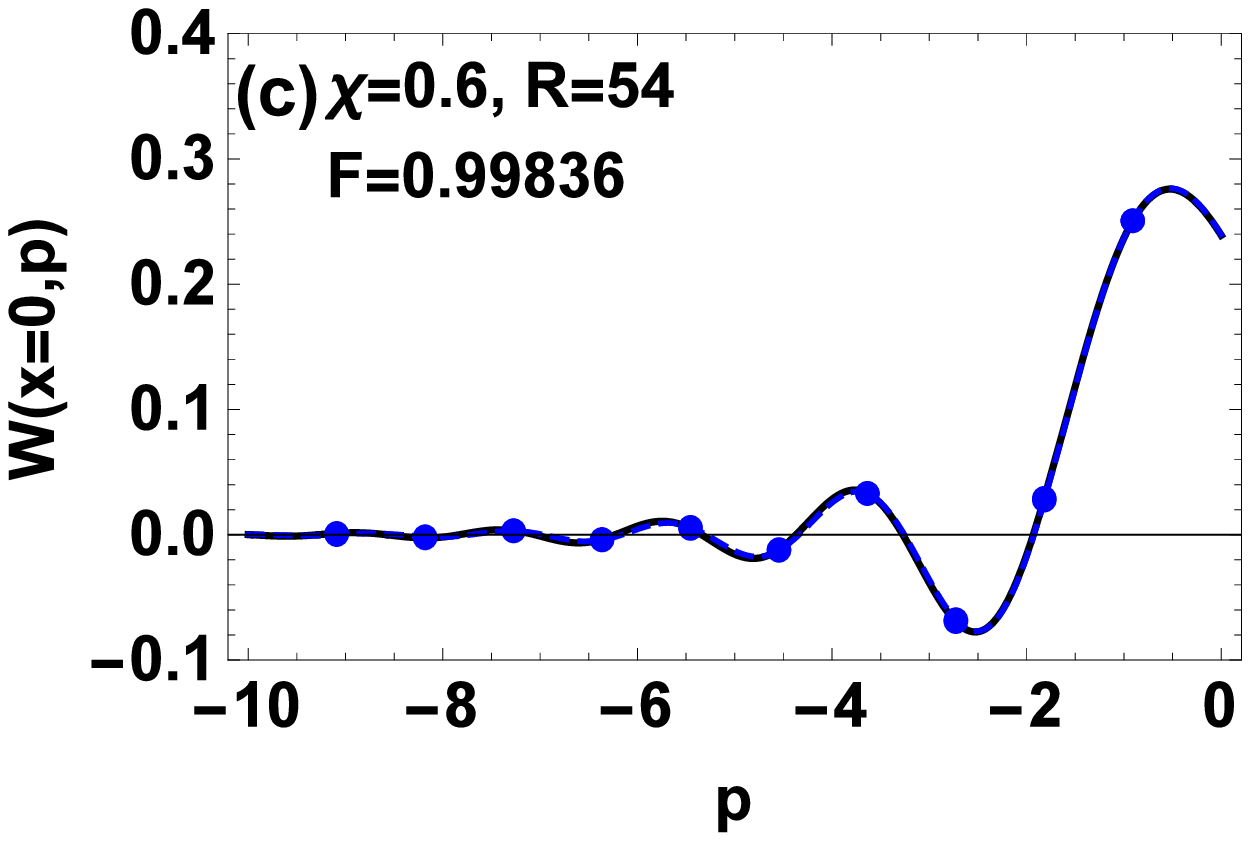}
\includegraphics[width=120px]{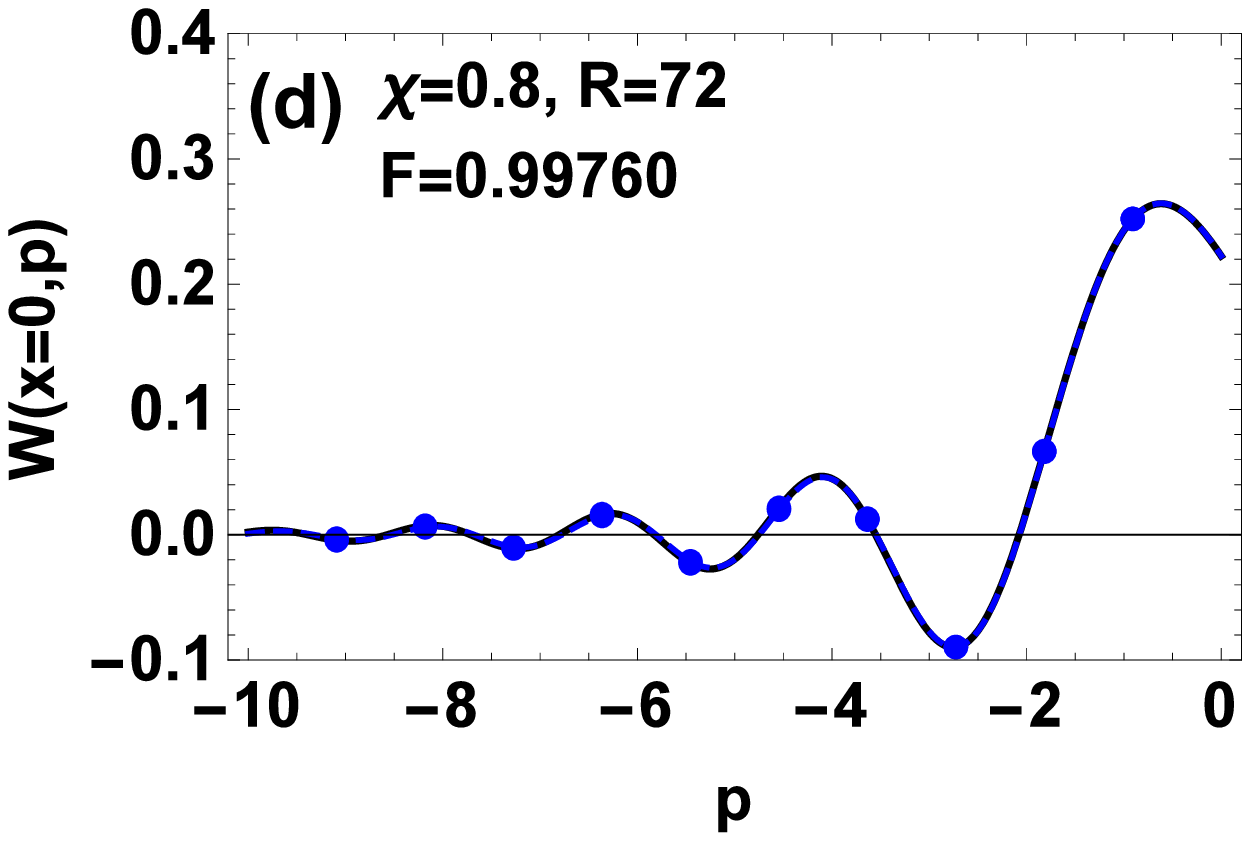}
\caption{ (Color online) Cross sections of Wigner functions at $x=0$ for $\chi=0.2, 0.4, 0.6, 0.8$ for a cubic gate. We observe a nice agreement between the overlapping ideal and generated Wigner functions.  Repition numbers of single rounds are $R=18, 36, 54, 72$, respectively. Fidelities with the target states are shown as insets. }
\label{wigner}
\end{figure}

In addition to quantitative figures of merit, we can also look for qualitative ones to check the nonclassical aspects of the engineered states. Wigner functions describe quantum states in quadrature phase space analogous to the classical probability densities. Nonclassical states produced by quantum nonlinearities exhibit peculiar properties, one of which is the presence of negative values. Each nonclassical state has a specific pattern of these nonclassical regions, and we can therefore check whether the states generated by the approximate cubic gate exhibit analogous patterns as the ideal cases. When the vacuum is the initial state, the ideal cubic states $\hat{U}_3\ket{0}$ exhibit negative fringes in the area given by $p<0$ around $x=0$, where $x, p$ are the eigenvalues of the quadrature operators $\hat{X}$ and $\hat{P} = (\hat{a} -\hat{a}^{\dag})/\sqrt{2}i$. In Fig~\ref{wigner}, we have plotted Wigner function values $W(x=0,p)$ for several cubic states at different strengths and the corresponding number of repetitions. We see that the two Wigner functions overlap nearly indistinguishably, and the approximate gate closely follows the ideal scenario, which confirms the gradual reproduction of the nonlinear dynamics.

 \begin{figure}[th]
\includegraphics[width=120px]{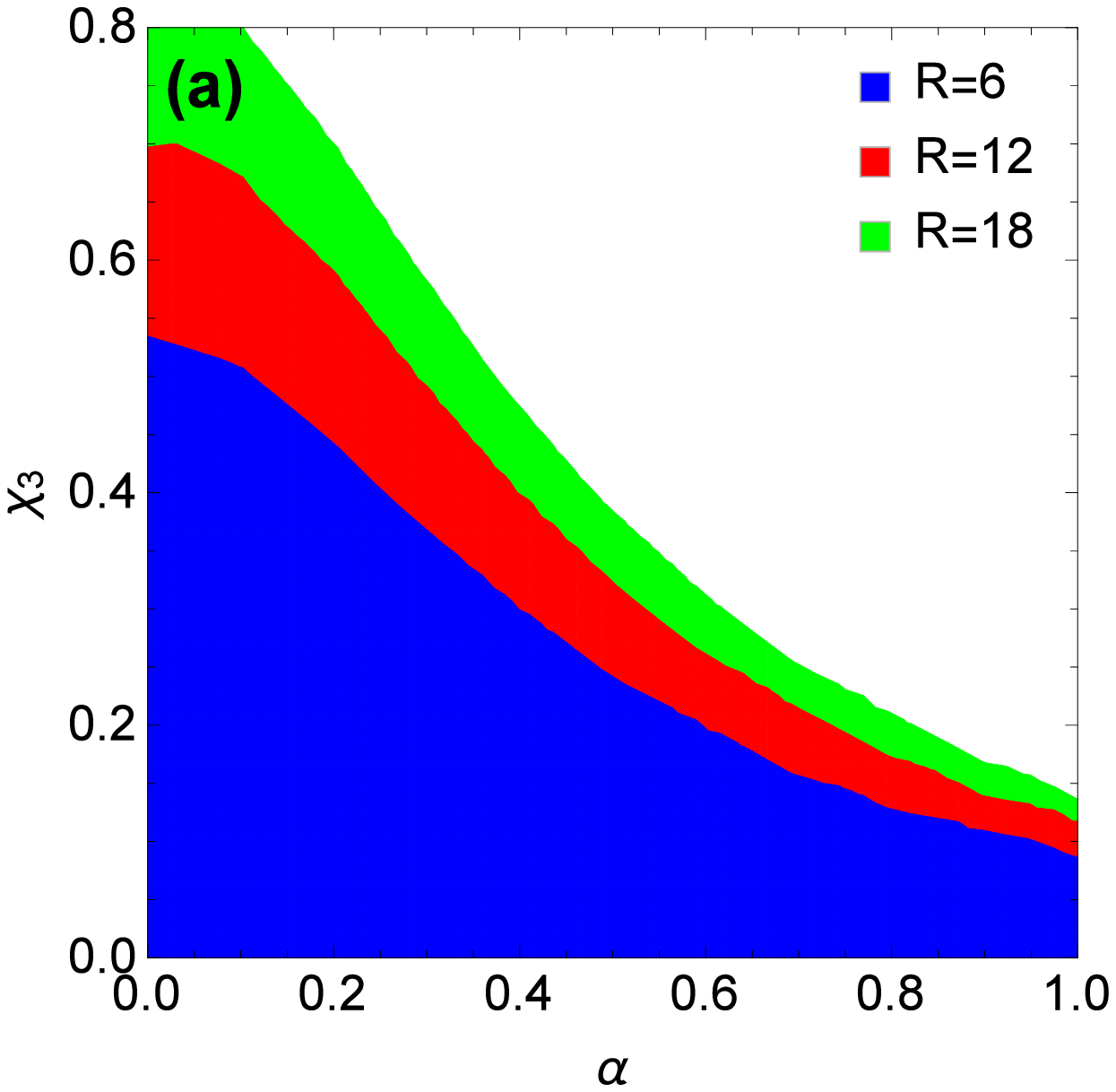}
\includegraphics[width=120px]{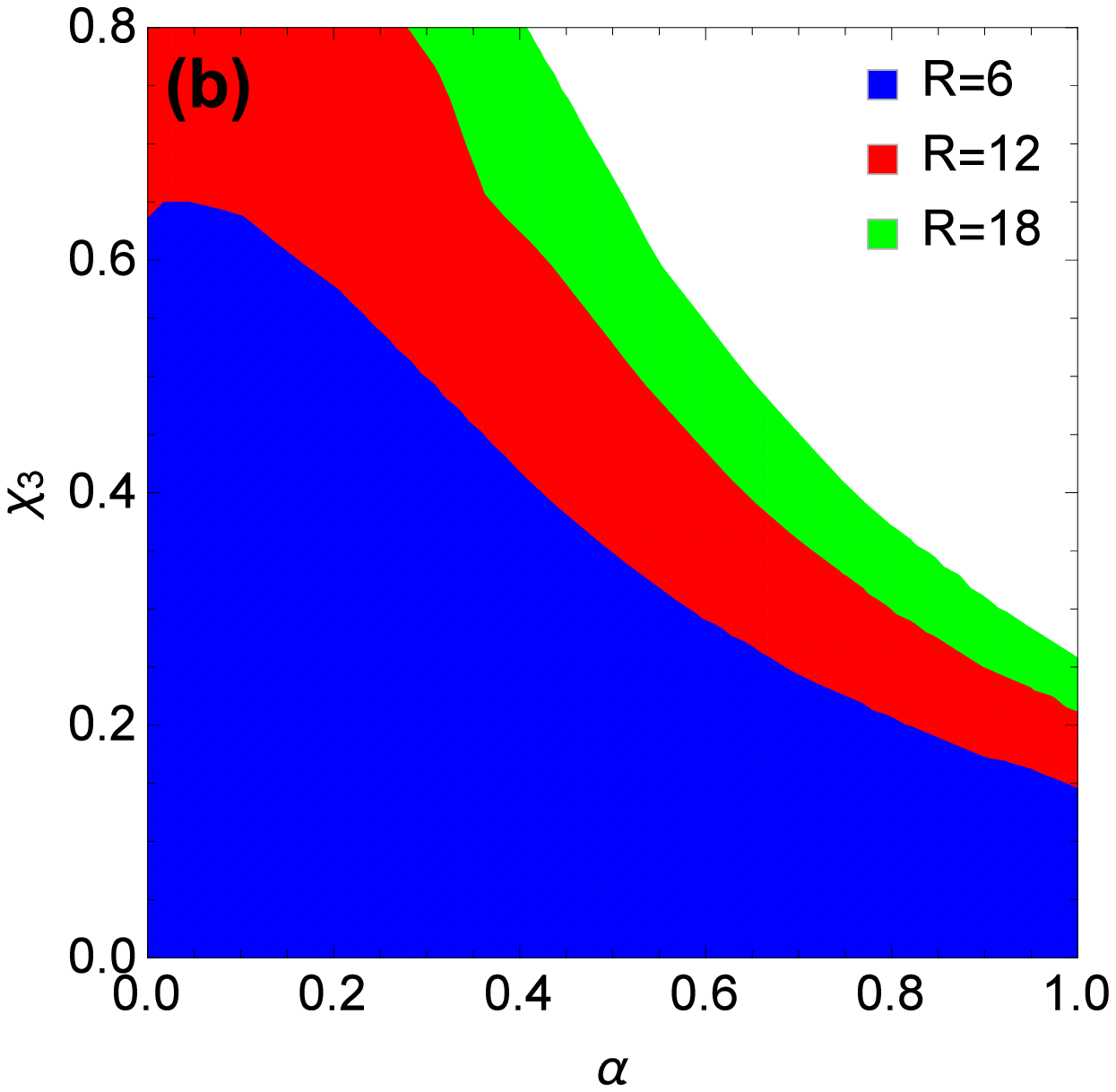}
\caption{(a) The region for which fidelity to the ideal state is larger than $0.99$ for a cubic phase gate. With a relatively low number of repetition, a high strength cubic gate is achieved for broad amplitudes of coherent states. The area of the regions can be expanded to an indefinite $\chi_3\rightarrow \infty$ when a high resource is accessible $R\rightarrow \infty$. (b) The region for which the fidelity of change $F_\mathrm{\perp}$ to the ideal state is larger than $0.95$ for the same generated operators. The fidelity of change is smaller than the full fidelity, but we still can observe the expansion when a larger $R$ is used.}
\label{region}
\end{figure}

To test the validity of our scheme for arbitrary input states, we have considered a set of coherent states $\ket{\alpha}=\exp(-\sqrt{2} i \alpha \hat{P})\ket{0}$ with $\alpha>0$ representing the amplitude of the state. Coherent states are suitable for feasible verification  due to easy generation in an experimental setting, while spanning the entire Hilbert space to form an overcomplete basis.  Fig.~\ref{region} (a) shows the region over $\alpha$ and $\chi$ where the fidelity with the ideal cubic gate is larger than a chosen threshold $F>0.99$. We can see that even with a relatively low number of repetitions $R$, a high strength cubic gate is achieved at a reasonable fidelity for broad range of amplitudes of coherent states, while a higher number of rounds is necessary for the coherent states to achieve a same fidelity for the vacuum state. Fig.~\ref{region} (b) shows, analogously, the area in which the fidelity of change satisfies $F_{\perp}>0.95$. Both of these results imply that the approximate cubic gate scheme is applicable to a wide class of states and that increasing the number of repetitions can extend this range.

 \begin{figure}[thp]
\includegraphics[width=120px]{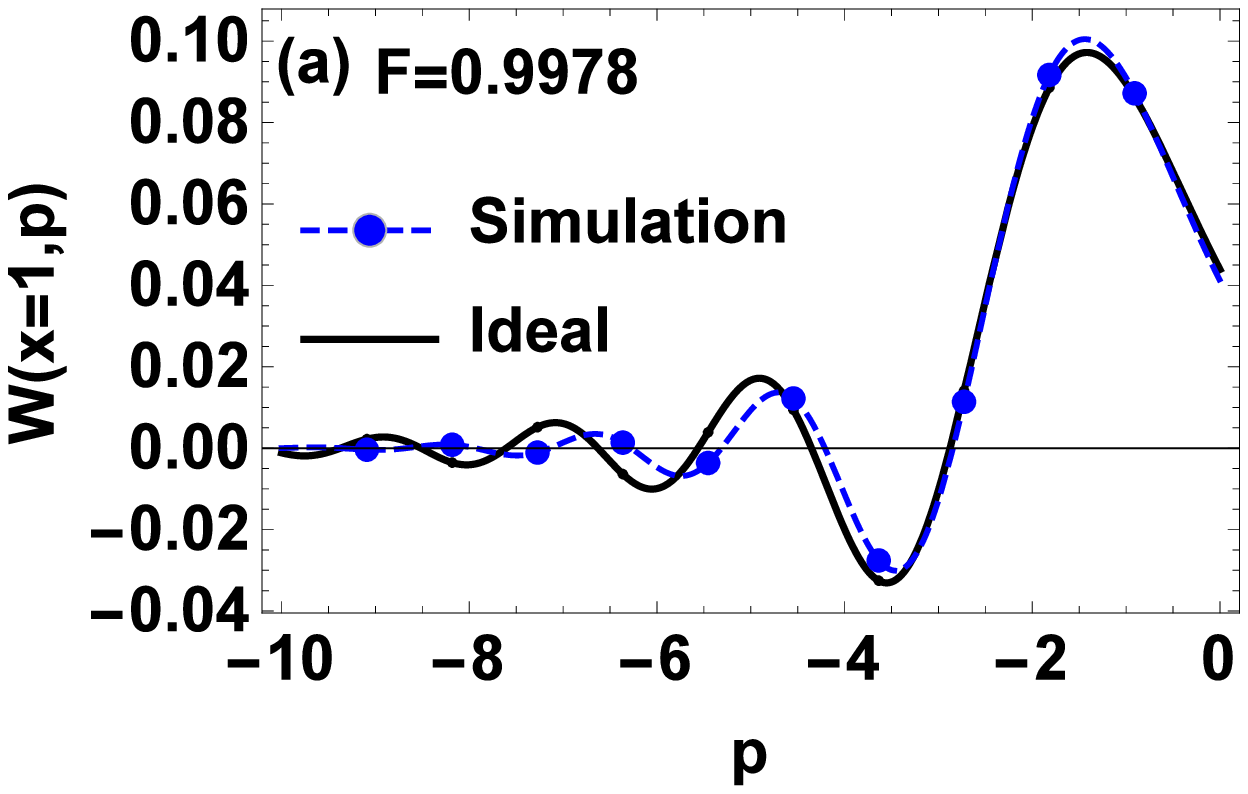}
\includegraphics[width=120px]{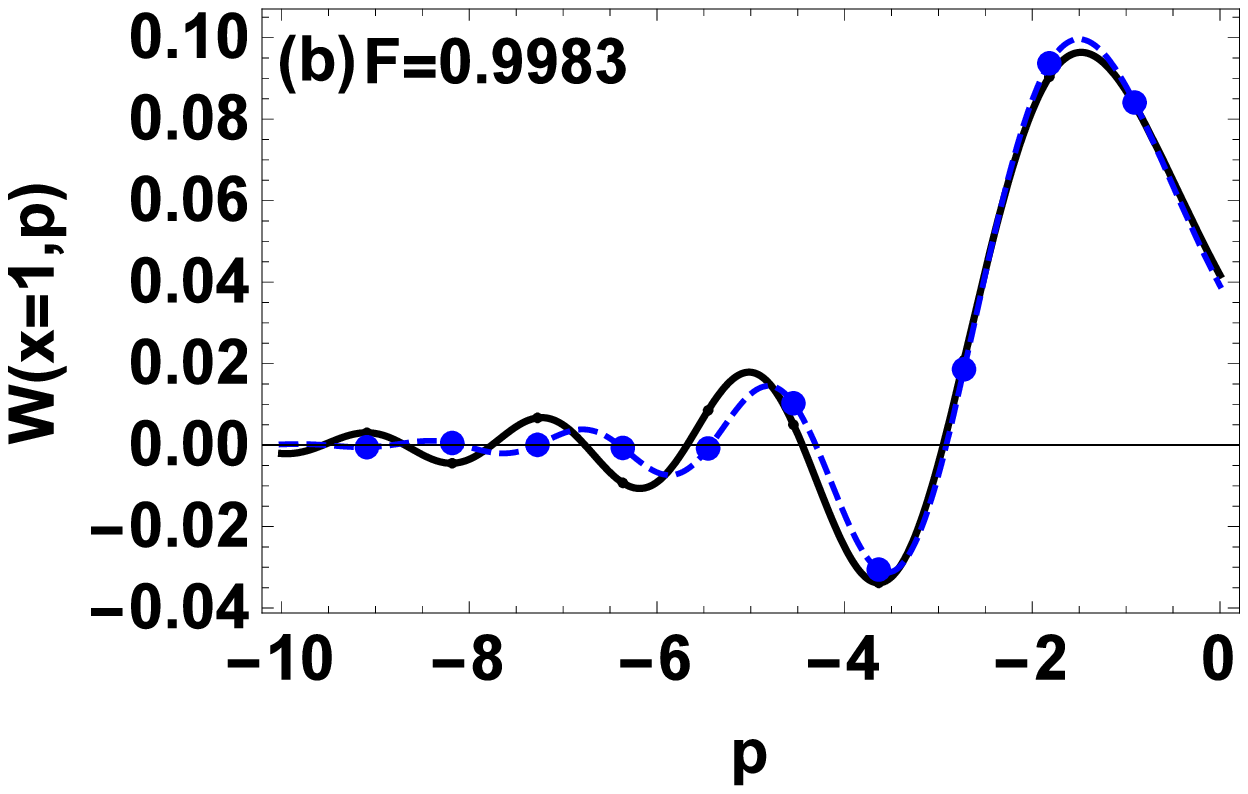}
\includegraphics[width=120px]{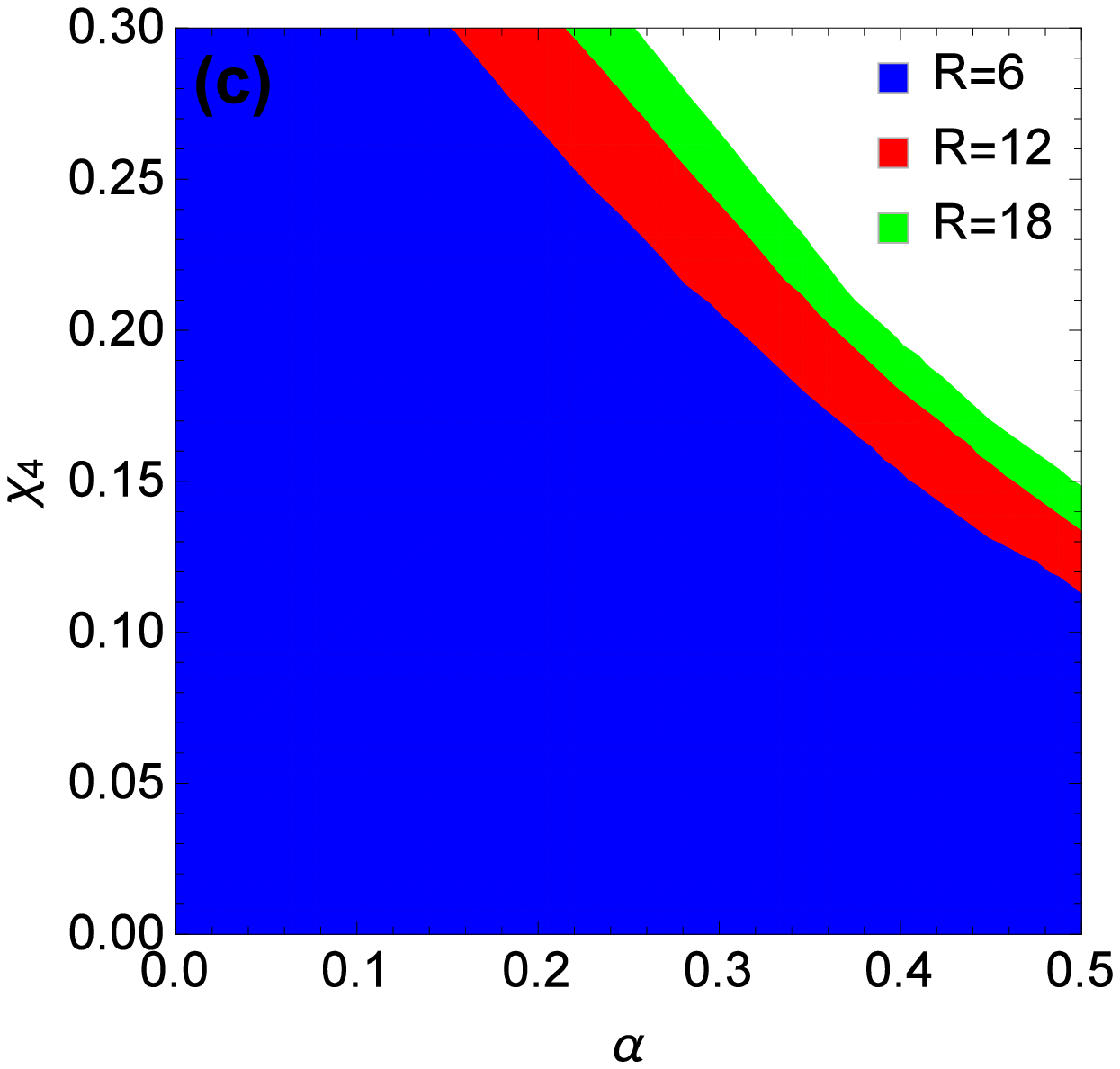}
\includegraphics[width=120px]{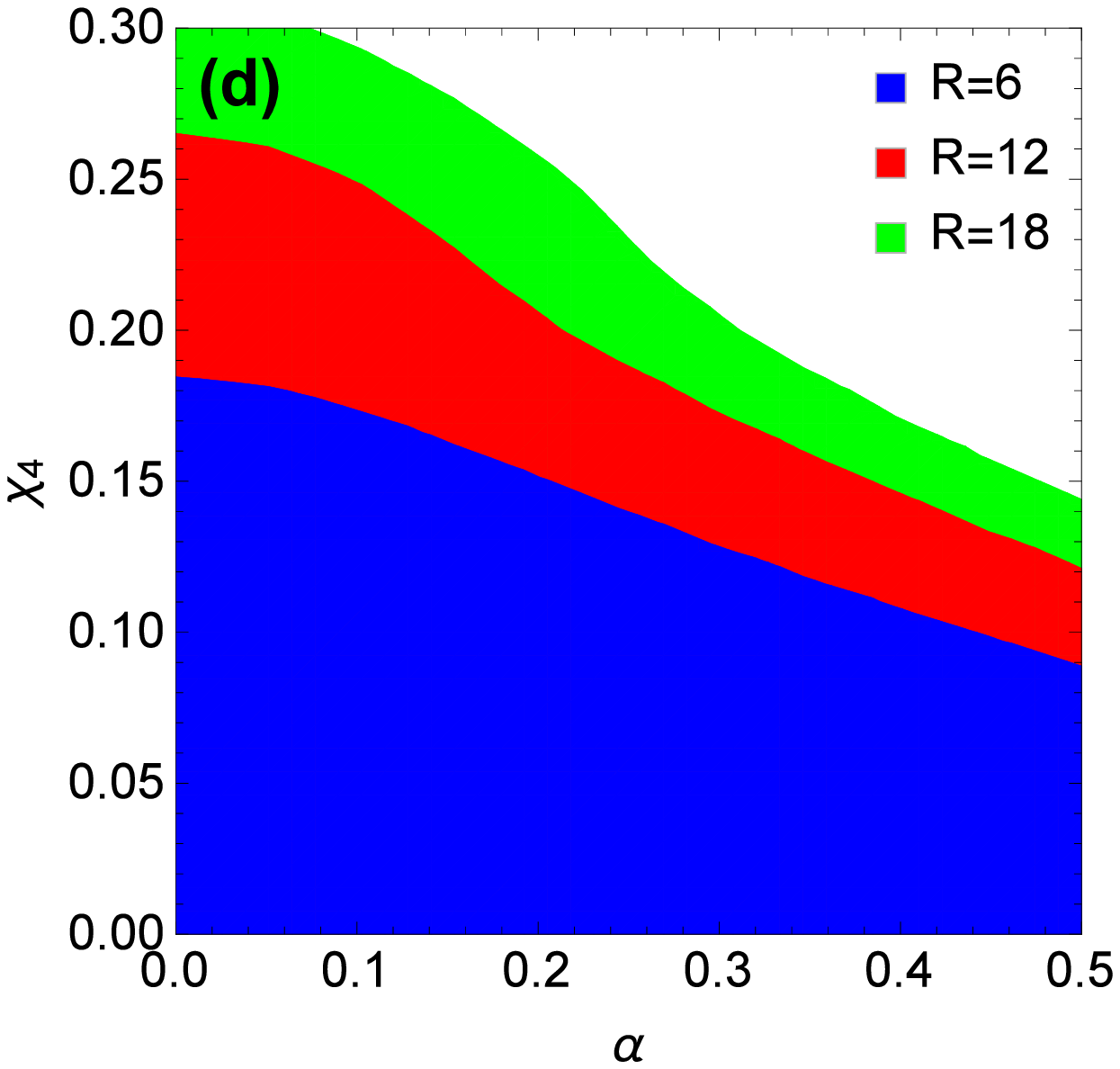}
\caption{The cross section of the Wigner function at $x=1$ after a quartic gate of (a) $\chi_4=0.2$ and (b) $\chi_4=0.4$. We notice that the overall oscillation is qualitatively analogous. The region for which (c) $F>0.99$ and (d) $F_\perp>0.95$ for a quartic gate. The area of the regions can again be expanded to an indefinite $\chi$ when a high $R$ is accessible.}
\label{quartregion}
\end{figure}

The basic principles applied to the cubic gate can be extended towards the fourth- and higher-order nonlinear quadrature phase gates. The fourth order quartic gate $\hat{U}_4=\exp[i\chi_4\hat{X}^4]$ can be realized by a sequence of elementary Rabi interactions
\begin{align}
&\hat{O}_{4,s}=\bra{g}\hat{M}_2(t_1 \hat{X},t_2 \hat{X})\hat{M}_2(-t_1 \hat{X},-t_2 \hat{X})\ket{g}\nonumber\\
&\approx\exp[i2t_2\hat{X}\sin[2t_1 \hat{X}]]
\label{eq:5-query}.
\end{align}
As in the case of the cubic gate, the crucial step lies in compensation of the lowest order term $\exp[i4 t_2t_1 \hat{X}^2]$ in the approximate expansion $\exp[i2t_2\hat{X}\left(2t_1 \hat{X}-(2t_1 \hat{X})^3/3!\right)]$ for small values of $t_1$. In this case, the correction step needs to be realized by a squeezing operation $G_c=\exp[-i4\zeta t_2t_1 \hat{X}^2]$. This squeezing operation can be similarly engineered from non-commuting Rabi interactions from $\hat{O}_{4,s}$ with different parameters $t_1'$ and $t'_2$. The squeezing parameter $\zeta$ can be optimized for the best performance. 
In Fig.~\ref{quartregion}, we perform the same analysis for Wigner functions, $F$ and  $F_\perp$ as for cubic gates. The simulation of the negative regions in the Wigner function of the quartic phase gate again shows a qualitative similarity, with only a small deviation. We can also achieve interaction with an arbitrary strength $\chi_4$, even though the requirements are stricter than for the cubic gate.

In general, the phase gates $\hat{U}_m=\exp[i\chi_m \hat{X}^m]$ for an arbitrary integer $m$ can be similarly achieved by applying (\ref{eq:5-query-c}) when $m$ is odd, or (\ref{eq:5-query}) when  $m$ is even, and eliminating all the undesirable lower order terms by already acheived phase gates $G_c=\hat{U}_{k<m}$. The proposed deterministic scheme  asymptotically realize an arbitrary unitary nonlinear quadrature phase gate achievable with the current technology. 
This scheme requires a sequence of controllable Rabi interactions between a harmonic oscillator and an ancillary two-level system~\cite{RabiOperationIon, RabiOperationCQED}, and the ability to periodically re-initialize the two-level system in its ground state. This sequence needs to be performed on timescales faster than the free evolution of the systems. Each individual sequence of Rabi interactions realizes a weak nonlinear quadrature phase gate, and stronger gates can be obtained by repeating the elementary steps. This repetition allows building up the desired operations incrementally and monitoring the dynamic evolution of the systems. This approach allows closing the universal set of operators in continuous variable quantum computation.

\begin{acknowledgments}
We acknowledge Project GB14-36681G of the Czech Science Foundation. K.P. acknowledges support by the Development Project of Faculty of Science, Palack\'y University. Authors thank L. Slodi\v{c}ka for a helpful discussion.
\end{acknowledgments}




\begin{thebibliography}{99}
\bibitem{CVqinf} C. Weedbrook, S. Pirandola, R. C.-Patr\'{o}n, N. J. Cerf, T. C. Ralph, J. H. Shapiro, and S. Lloyd, Rev. Mod. Phys. 84, 621 (2012); S. L. Braunstein, and P. Van Loock,  Rev. Mod. Phys., 77(2), 513 (2005);  N. J. Cerf, G. Leuchs, and E. S. Polzik (eds.), \textit{Quantum
Information with Continuous Variables of Atoms and Light}
(Imperial College Press, London, 2007).
\bibitem{CVlight} P. A. M. Dirac, \textit{The Quantum Theory of the Emission and Absorption of Radiation}, Proc. Royal Soc. (London) A114, pp. 243–265 (1927).
\bibitem{CVatoms} T. Opatrny, arXiv:1702.03124 (2017); I. D. Leroux, M. H. Schleier-Smith, and V. Vulet\'ic,  Phys. Rev. Lett. 104, 073602 (2010).
\bibitem{CVions} D. Leibfried, R. Blatt, C. Monroe, and D. Wineland, Rev. Mod. Phys. 75, 281 (2003).
\bibitem{CVcavQED} A. Reiserer and G. Rempe, Rev. Mod. Phys. 87, 1379 (2015).
\bibitem{CVcirQED} 
A. Blais, R.-S. Huang, A. Wallraff, S. M. Girvin, and R. J. Schoelkopf, Phys. Rev. A 69, 062320 (2004); A. Wallraff, D. I. Schuster, A. Blais, L. Frunzio, R.- S. Huang, J. Majer, S. Kumar, S. M. Girvin and R. J. Schoelkopf, Nature 431, 162 (2004). \bibitem{LloydPRL1999Universal} S. Lloyd and S. L. Braunstein, Phys. Rev. Lett. 82, 1784 (1999).
\bibitem{cubicTrapped} M. \v{S}iler, P. J\'akl, O. Brzobohat\'y, A. Ryabov, R. Filip, and P. Zem\'anek,  Sci. Rep. 7, 1697 (2017).
\bibitem{nonlinearity}  L. Lugiato, F. Prati, and M. Brambilla, \textit{Nonlinear Optical
Systems} (Cambridge University Press, Cambridge, 2015).

\bibitem{ParkPRA2016Rabiconditional} K. Park, P. Marek and R. Filip, Phys. Rev. A 94, 062308 (2016).
\bibitem{ParkPRA2016JC} K. Park, P. Marek and R. Filip, A 94, 012332 (2016).\bibitem{OptCub}  S. Sefi and P. van Loock, Phys. Rev. Lett. 107, 170501 (2011); F. Arzani, N. Treps, G. Ferrini, Phys. Rev. A 95, 052352 (2017).

\bibitem{cubicstate} D. Gottesman, A. Kitaev, and J. Preskill, Phys. Rev. A 64,
012310 (2001); S. D. Bartlett and B. C. Sanders, Phys. Rev. A 65,042304 (2002).
\bibitem{BartlettPRA2002cubicstate}  S. D. Bartlett and B. C. Sanders, Phys. Rev. A 65, 042304 (2002).
\bibitem{MarekPRA2011Cubic} P. Marek, R. Filip, and A. Furusawa, Phys. Rev. A 84, 053802 (2011).
\bibitem{MiyataPRA2016Cubic} K. Miyata, H. Ogawa, P. Marek, R. Filip, H. Yonezawa, J.-i. Yoshikawa, and A. Furusawa, Phys. Rev. A 93, 022301 (2016).




\bibitem{ParkKerrSciRep2017} K. Park, P. Marek and R. Filip, Deterministic nonlinear gates with oscillators mediated by a qubit, arXiv: 1706.09020 (2017).
\bibitem{KrastanovConstructivePRA2015}  S. Krastanov, V. V. Albert, C. Shen, C.-L. Zou, R. W. Heeres, B. Vlastakis, R. J. Schoelkopf, and L. Jiang, Phys. Rev. A 92, 040303 (R) (2015);
W. Heeres, B. Vlastakis, E. Holland, S. Krastanov, V. V. Albert, L. Frunzio, L. Jiang, and R. J. Schoelkopf, Phys. Rev. Lett. 115, 137002 (2015).
\bibitem{Nonclassicality}
P. Marek, L. Lachman, L. Slodi\v{c}ka, and R. Filip, Phys. Rev. A 94, 013850 (2016);
L. Slodi\v{c}ka,. P. Marek, and R. Filip, Opt. Exp. 24, 7858 (2016).
\bibitem{simulation} I. M. Georgescu,  S. Ashhab, and F. Nori, Rev. Mod. Phys. 86, 153 (2014).

\bibitem{RabiOperationIon} H.-Y. Lo,	D. Kienzler,	L. de Clercq,	M. Marinelli,	V. Negnevitsky,	B. C. Keitch,	and J. P. Home, Nature 521, 336–339 (2015); D. Kienzler, C. Fl\"uhmann, V. Negnevitsky, H.-Y. Lo, M. Marinelli, D.
Nadlinger, and J. P. Home, Phys. Rev. Lett. 116, 140402 (2016); J. S. Pedernales,  I. Lizuain, S. Felicetti, G. Romero, L. Lamata, and E. Solano, Sci. Rep. 5, 15472 (2015).
\bibitem{RabiOperationCQED} T. Niemczyk, F. Deppe, H. Huebl, E. P. Menzel, F. Hocke, M. J. Schwarz, J. J. Garcia-Ripoll,
D. Zueco, T. H{\"u}mmer, E. Solano, A. Marx and R. Gross, Nat. Phys. 6, 772 (2010); A. Mezzacapo, U. Las Heras, J. S. Pedernales, L. DiCarlo, E. Solano and L. Lamata,  Sci. Rep. 4, 7482 (2013); M. A. Sillanp\"{a}\"{a}, J. I. Park, and R. W. Simmonds, Nature (London) 449, 438 (2007); G. G\"{u}nter, A. A. Anappara, J. Hees, A. Sell, G. Biasiol, L. Sorba, S. De Liberato, C. Ciuti, A. Tredicucci, A. Leitenstorfer, and R. Huber, ibid. 458,178(2009); P. Forn-D\'iaz, J. Lisenfeld, D. Marcos, J. J.
Garc\'ia-Ripoll, E. Solano, C. J. P. M. Harmans, and J. E. Mooij, Phys. Rev. Lett.
105, 237001 (2010); A. Baust, E. Hoffmann, M. Haeberlein, M.
J. Schwarz, P. Eder, J. Goetz, F. Wulschner, E. Xie, L. Zhong,
F. Quijandr\'ia, D. Zueco, J.-J. Garc\'ia Ripoll, L. Garc\'ia-\'Alvarez,
G. Romero, E. Solano, K. G. Fedorov, E. P. Menzel, F. Deppe,
A. Marx, and R. Gross, Phys. Rev. B 93, 214501 (2016).
\bibitem{RabiHist}  I. I. Rabi, Phys. Rev.
49, 324 (1936); ibid. 51, 652 (1937); J. H. Eberly, N. B. Narozhny, and J. J. Sanchez-Mondragon, Phys. Rev. Lett. 44, 1323 (1980).
\bibitem{BCH} R. Achilles and A. Bonfiglioli,  Arch. Hist. Exact Sci. 66, 295 (2012). 
\bibitem{CQEDtheory}  Z. Y. Xue, Y. F. Li, J. Zhou, Y. M. Gao, and G. Zhang,    Quantum Inf. Process. 15, 721 (2016); T. Lindstr\"om, C. H. Webster, J. E. Healey, M. S. Colclough, C. M. Muirhead, and A. Y. Tzalenchuk, Supercond. Sci. Tech.
20, 814 (2007).



 




\end{thebibliography}
\end{document}